%#!tex CSmatrix.tex
%% Last Modified: Wed Jun  7 17:15:15 2006.

%%%%%%%%%%%%%%%%%%%%%%%%%%%%%%%%%%%%%%%%%%%%%%%%%%%%%%%%%%%%%%%%%
%                                                               %
%                          NS5                                  %
%                                                               %
%                  Kazumi Okuyama (UBC)                         %
%                                                               %
%%%%%%%%%%%%%%%%%%%%%%%%%%%%%%%%%%%%%%%%%%%%%%%%%%%%%%%%%%%%%%%%%

%\input harvmac

%\def\mydraft{label}

\input lanlmac
\input amssym
\input epsf

%Macro for figure
\newcount\figno
\figno=0
\def\fig#1#2#3{
\par\begingroup\parindent=0pt\leftskip=1cm\rightskip=1cm\parindent=0pt
\baselineskip=13pt
\global\advance\figno by 1
\midinsert
\epsfxsize=#3
\centerline{\epsfbox{#2}}
\vskip 12pt
%\centerline{{\bf Fig. \the\figno:~~} #1}\par
{\bf Fig. \the\figno:~~} #1 \par
\endinsert\endgroup\par
}
\def\figlabel#1{\xdef#1{\the\figno}}
%%                              TABLEAUX.TEX
%%      This  macro file is for producing a ``Young Tableau'' which is
%%      an array of little squares sometimes used in mathematical physics.
%%      For instance, the command $\tableau{6 3 2}$ will produce a tableau
%%      with 6 squares in the top row, 3 in the next, and 2 in the last.
%%                                  OOOOOO
%%      This tableau will look like OOO    but made of squares instead of
%%           O's.
%%                                  OO
%%      Any number of rows may be present, each having a nonzero number of
%%      squares.
%%
%%      A tableau is math mode material, so use $ or $$ to enclose it.
%%
%%      The size and line-thickness of the little boxes are controlled by
%%  the
%%      dimension parameters --
%%              \tableauside=1.0ex              %(size)
%%              \tableaurule=0.4pt              %(line-thickness)
%%      Change them if you want.
%%
%%                                                      -- Doug Eardley
%%   9/19/8
%%
%%
\newdimen\tableauside\tableauside=1.0ex
\newdimen\tableaurule\tableaurule=0.4pt
\newdimen\tableaustep
\def\phantomhrule#1{\hbox{\vbox to0pt{\hrule height\tableaurule
width#1\vss}}}
\def\phantomvrule#1{\vbox{\hbox to0pt{\vrule width\tableaurule
height#1\hss}}}
\def\sqr{\vbox{%
  \phantomhrule\tableaustep

\hbox{\phantomvrule\tableaustep\kern\tableaustep\phantomvrule\tableaustep}%
  \hbox{\vbox{\phantomhrule\tableauside}\kern-\tableaurule}}}
\def\squares#1{\hbox{\count0=#1\noindent\loop\sqr
  \advance\count0 by-1 \ifnum\count0>0\repeat}}
\def\tableau#1{\vcenter{\offinterlineskip
  \tableaustep=\tableauside\advance\tableaustep by-\tableaurule
  \kern\normallineskip\hbox
    {\kern\normallineskip\vbox
      {\gettableau#1 0 }%
     \kern\normallineskip\kern\tableaurule}% 
  \kern\normallineskip\kern\tableaurule}}
\def\gettableau#1 {\ifnum#1=0\let\next=\null\else
  \squares{#1}\let\next=\gettableau\fi\next}

\tableauside=1.0ex
\tableaurule=0.4pt

%Macro

\def\CS{{\cal S}}
\def\CZ{{\cal Z}}

\def\Tr{{\rm Tr}}
\def\hf{{1\over 2}}

\def\o{\over}

\def\b#1{\overline{#1}}
\def\del{\partial}

\def\lap{\Delta}
\def\bra{\langle}
\def\ket{\rangle}
\def\lf{\left}
\def\ri{\right}
\def\riya{\rightarrow}

\def\al{\alpha}

\def\rt#1{\sqrt{#1}}

\def\sitarel#1#2{\mathrel{\mathop{\kern0pt #1}\limits_{#2}}}

%\newsec{References}
\lref\MaldacenaSN{
  J.~M.~Maldacena, G.~W.~Moore, N.~Seiberg and D.~Shih,
  ``Exact vs. semiclassical target space of the minimal string,''
  JHEP {\bf 0410}, 020 (2004)
  [arXiv:hep-th/0408039].
  %%CITATION = HEP-TH 0408039;%%
}
\lref\MorozovHH{
  A.~Morozov,
  ``Integrability and matrix models,''
  Phys.\ Usp.\  {\bf 37}, 1 (1994)
  [arXiv:hep-th/9303139].
  %%CITATION = HEP-TH 9303139;%%
}
\lref\OkudaMB{
  T.~Okuda,
  ``Derivation of Calabi-Yau crystals from Chern-Simons gauge theory,''
  JHEP {\bf 0503}, 047 (2005)
  [arXiv:hep-th/0409270].
  %%CITATION = HEP-TH 0409270;%%
}
\lref\IqbalDS{
  A.~Iqbal, N.~Nekrasov, A.~Okounkov and C.~Vafa,
  ``Quantum foam and topological strings,''
  arXiv:hep-th/0312022.
  %%CITATION = HEP-TH 0312022;%%
}
\lref\OkounkovSP{
  A.~Okounkov, N.~Reshetikhin and C.~Vafa,
  ``Quantum Calabi-Yau and classical crystals,''
  arXiv:hep-th/0309208.
  %%CITATION = HEP-TH 0309208;%%
}
\lref\SaulinaDA{
  N.~Saulina and C.~Vafa,
  ``D-branes as defects in the Calabi-Yau crystal,''
  arXiv:hep-th/0404246.
  %%CITATION = HEP-TH 0404246;%%
}
\lref\OoguriBV{
  H.~Ooguri and C.~Vafa,
  ``Knot invariants and topological strings,''
  Nucl.\ Phys.\ B {\bf 577}, 419 (2000)
  [arXiv:hep-th/9912123].
  %%CITATION = HEP-TH 9912123;%%
}
\lref\HalmagyiVK{
  N.~Halmagyi, A.~Sinkovics and P.~Sulkowski,
  ``Knot invariants and Calabi-Yau crystals,''
  JHEP {\bf 0601}, 040 (2006)
  [arXiv:hep-th/0506230].
  %%CITATION = HEP-TH 0506230;%%
}
\lref\GaiottoYB{
  D.~Gaiotto and L.~Rastelli,
  ``A paradigm of open/closed duality: Liouville D-branes and the  Kontsevich
  model,''
  JHEP {\bf 0507}, 053 (2005)
  [arXiv:hep-th/0312196].
  %%CITATION = HEP-TH 0312196;%%
}
\lref\MarinoFK{
  M.~Marino,
  ``Chern-Simons theory, matrix integrals, and perturbative three-manifold
  invariants,''
  Commun.\ Math.\ Phys.\  {\bf 253}, 25 (2004)
  [arXiv:hep-th/0207096].
  %%CITATION = HEP-TH 0207096;%%
}
\lref\AganagicWV{
  M.~Aganagic, A.~Klemm, M.~Marino and C.~Vafa,
  ``Matrix model as a mirror of Chern-Simons theory,''
  JHEP {\bf 0402}, 010 (2004)
  [arXiv:hep-th/0211098].
  %%CITATION = HEP-TH 0211098;%%
}
\lref\AganagicQJ{
  M.~Aganagic, R.~Dijkgraaf, A.~Klemm, M.~Marino and C.~Vafa,
  ``Topological strings and integrable hierarchies,''
  Commun.\ Math.\ Phys.\  {\bf 261}, 451 (2006)
  [arXiv:hep-th/0312085].
  %%CITATION = HEP-TH 0312085;%%
}
\lref\DijkgraafVP{
  R.~Dijkgraaf, A.~Sinkovics and M.~Temurhan,
  ``Universal correlators from geometry,''
  JHEP {\bf 0411}, 012 (2004)
  [arXiv:hep-th/0406247].
  %%CITATION = HEP-TH 0406247;%%
}
\lref\GopakumarVY{
  R.~Gopakumar and C.~Vafa,
  ``Topological gravity as large N topological gauge theory,''
  Adv.\ Theor.\ Math.\ Phys.\  {\bf 2}, 413 (1998)
  [arXiv:hep-th/9802016].
  %%CITATION = HEP-TH 9802016;%%
}
\lref\GopakumarKI{
  R.~Gopakumar and C.~Vafa,
  ``On the gauge theory/geometry correspondence,''
  Adv.\ Theor.\ Math.\ Phys.\  {\bf 3}, 1415 (1999)
  [arXiv:hep-th/9811131].
  %%CITATION = HEP-TH 9811131;%%
}
\lref\TierzJJ{
  M.~Tierz,
  ``Soft matrix models and Chern-Simons partition functions,''
  Mod.\ Phys.\ Lett.\ A {\bf 19}, 1365 (2004)
  [arXiv:hep-th/0212128].
  %%CITATION = HEP-TH 0212128;%%
}
\lref\deHaroRZ{
  S.~de Haro and M.~Tierz,
  ``Discrete and oscillatory matrix models in Chern-Simons theory,''
  Nucl.\ Phys.\ B {\bf 731}, 225 (2005)
  [arXiv:hep-th/0501123].
  %%CITATION = HEP-TH 0501123;%%
}
\lref\Szego{
G. Szeg\"{o}, {\it Orthogonal Polynomials}, Colloquium Publications, Vol. 23
(American Mathematical Society).
}
\lref\FaddeevRS{
  L.~D.~Faddeev and R.~M.~Kashaev,
  ``Quantum Dilogarithm,''
  Mod.\ Phys.\ Lett.\ A {\bf 9}, 427 (1994)
  [arXiv:hep-th/9310070].
  %%CITATION = HEP-TH 9310070;%%
}
\lref\OoguriGX{
  H.~Ooguri and C.~Vafa,
  ``Worldsheet derivation of a large N duality,''
  Nucl.\ Phys.\ B {\bf 641}, 3 (2002)
  [arXiv:hep-th/0205297].
  %%CITATION = HEP-TH 0205297;%%
}
\lref\Walker{
P. J. Forrester, ``Vicious random walkers in the limit of a large number of walkers,'' J. Stat. Phys. {\bf 56}, 767 (1989).
}
\lref\WittenFB{
  E.~Witten,
  ``Chern-Simons gauge theory as a string theory,''
  Prog.\ Math.\  {\bf 133}, 637 (1995)
  [arXiv:hep-th/9207094].
  %%CITATION = HEP-TH 9207094;%%
}
\lref\WittenHF{
  E.~Witten,
  ``Quantum Field Theory And The Jones Polynomial,''
  Commun.\ Math.\ Phys.\  {\bf 121}, 351 (1989).
  %%CITATION = CMPHA,121,351;%%
}
\lref\BershadskyCX{
  M.~Bershadsky, S.~Cecotti, H.~Ooguri and C.~Vafa,
  ``Kodaira-Spencer theory of gravity and exact results for quantum string
  amplitudes,''
  Commun.\ Math.\ Phys.\  {\bf 165}, 311 (1994)
  [arXiv:hep-th/9309140].
  %%CITATION = HEP-TH 9309140;%%
}
\lref\HoriKT{
  K.~Hori and C.~Vafa,
  ``Mirror symmetry,''
  arXiv:hep-th/0002222.
  %%CITATION = HEP-TH 0002222;%%
}
\lref\MarinoEQ{
  M.~Marino,
  ``Les Houches lectures on matrix models and topological strings,''
  arXiv:hep-th/0410165.
  %%CITATION = HEP-TH 0410165;%%
}
\lref\Andrews{
C. E. Andrews, {\it The Theory of Partitions} (Cambridge University Press,
1998).
}
\lref\AganagicGS{
  M.~Aganagic and C.~Vafa,
  ``Mirror symmetry, D-branes and counting holomorphic discs,''
  arXiv:hep-th/0012041.
  %%CITATION = HEP-TH 0012041;%%
}
\lref\AganagicNX{
  M.~Aganagic, A.~Klemm and C.~Vafa,
  ``Disk instantons, mirror symmetry and the duality web,''
  Z.\ Naturforsch.\ A {\bf 57}, 1 (2002)
  [arXiv:hep-th/0105045].
  %%CITATION = HEP-TH 0105045;%%
}

%%%%%%%%%%%%%%%%%%%%%%%%%%%%%%%%%%%%%%%%%%%%%%%%%%%%%%%%%%%%%%%%%
%                      Title Page                               %
%%%%%%%%%%%%%%%%%%%%%%%%%%%%%%%%%%%%%%%%%%%%%%%%%%%%%%%%%%%%%%%%%
\Title{             
                                             \vbox{
                                             \hbox{hep-th/0606048}}}
{\vbox{
\centerline{D-Brane Amplitudes in Topological String on Conifold}
}}

\vskip .2in

\centerline{Kazumi Okuyama}

\vskip .2in

%\vskip 2cm
\centerline{Department of Physics and Astronomy, 
University of British Columbia} 
\centerline{Vancouver, BC, V6T 1Z1, Canada}
\centerline{\tt kazumi@phas.ubc.ca}
\vskip 3cm
\noindent

%%abstract
We study the relation between two kinds of
topological amplitudes of non-compact D-branes
on conifold. In the A-model, D-branes are represented by
fermion operators in the melting crystal picture and  
the amplitudes are given by the quantum dilogarithm.
In the mirror B-model, D-branes correspond to the determinant operator
$\det(x-M)$ in 
the Chern-Simons matrix model and the amplitudes 
are given by the Stieltjes-Wigert polynomial.
We show that these two amplitudes are related
by a certain integral transformation.
We argue that this transformation represents the
deformation of closed string background due to
the presence of D-branes.

\Date{June 2006}

\vfill
\vfill

\newsec{Introduction}
The gauge/string duality is one of the most 
important aspects of string theory.
It appears that this duality is also essential
for the topological string theory.
In \refs{\GopakumarVY,\GopakumarKI}, 
it is realized that the 't Hooft expansion of
Chern-Simons theory on $S^3$ is exactly the same as the
topological A-model on the resolved conifold 
${\cal O}(-1)\oplus{\cal O}(-1)\riya{\Bbb P}^1$.
It was shown that the partition function 
of $SU(N)$ level $k$ Chern-Simons theory on $S^3$ \WittenHF
\eqn\Zalprod{
Z_{CS}={1\o(k+N)^{N\o2}}\prod_{\al>0}2\sin{\pi(\al,\rho)\o k+N}
}
agrees with the partition function of topological closed string
on local ${\Bbb P}^1$ \BershadskyCX
\eqn\topZ{
Z_{top}=\prod_{n=1}^\infty\lf({1-Qq^n\o1-q^n}\ri)^n
}
up to a non-perturbative factor
$\eta(q)^N$. In this correspondence, various parameters are related as
\eqn\gstring{\eqalign{
g_s&={2\pi i\o k+N}~,\quad t=g_sN ,\cr
q&=e^{-g_s}~,\quad Q=e^{-t},
}}
where $g_s$ is the string coupling and
the 't Hooft parameter $t$
is identified as the K\"{a}hler parameter of ${\Bbb P}^1$. 
This is an example of geometric transition where
D-branes wrapped around $S^3$ in the deformed conifold
\WittenFB\ is replaced by a $B$-field on
$S^2$ in the resolved conifold.
In \OoguriGX, this duality was derived by using a linear sigma model.

In this paper, we consider open string
amplitudes associated with non-compact D-branes in this background.
There are two ways to define this amplitude.
The first way is to introduce non-compact Lagrangian A-branes
in the deformed conifold $T^*S^3$. 
The amplitude of the A-branes is given by Wilson loops
in Chern-Simons theory \OoguriBV.
Later it was shown that the same amplitude is obtained
in the melting crystal picture \refs{\OkounkovSP,\IqbalDS}.
In this picture, the A-brane corresponds to a defect of crystal and 
it is represented by a certain fermion operator $\Psi_D(x)$ \SaulinaDA . 
The amplitude $\bra\Psi_D(x)\ket$ 
is given by the quantum dilogarithm 
\refs{\SaulinaDA\OkudaMB\DijkgraafVP{--}\HalmagyiVK}. 

The second way to define the D-brane amplitude is to use the
matrix model description of
Chern-Simons theory \refs{\MarinoFK\AganagicWV\TierzJJ{--}\MarinoEQ}.
A natural D-brane amplitude in the Chern-Simons matrix model is the
expectation value of determinant $\bra\det(x-M)\ket$.
Since this matrix model appears as the mirror B-model of conifold
\AganagicWV, the determinant $\det(x-M)$ 
represents a non-compact B-brane in the mirror description \AganagicQJ.
It is well known that $\bra\det(x-M)\ket$ is equal to the $N$-th
orthogonal polynomial $P_N(x)$.
For the Chern-Simons matrix model, the associated orthogonal
polynomial is known as the Stieltjes-Wigert polynomial \Szego.

It is natural to ask what is the relation between the A-brane amplitude
$\bra\Psi_D(x)\ket$ in the crystal picture and the
B-brane amplitude $\bra\det(x-M)\ket$ in the
Chern-Simons matrix model.
In this paper, we will show that they are related by a simple Gaussian
integral.

This paper is organized as follows.
In section 2, we review the Gaussian matrix model as a 
preliminary of the analysis of Chern-Simons matrix model.
In section 3, we find a simple relation between A-brane amplitudes
and B-brane amplitudes.
Section 4 is discussion.
\newsec{Gaussian Matrix Model: Review}
Before discussing the D-branes in topological string, let us first
review a simple example, namely FZZT-branes in Gaussian matrix model
\MaldacenaSN. Although this is a too trivial example,
we will see that  most of the properties found in
the Gaussian matrix model have close analogues
in the Chern-Simons matrix model.

The Gaussian matrix model is defined by
\eqn\ZG{
Z_G=\int_{N\times N} dM e^{-{1\o2g_s}\Tr M^2}.
}
We are interested in the expectation value of the determinant
operator $\det(x-M)$
\eqn\detinG{
\bra\det(x-M)\ket={1\o Z_G}\int dMe^{-{1\o2g_s}\Tr M^2}\det(x-M),
}
which can be interpreted as the wavefunction of FZZT-brane.
As discussed in detail in \MaldacenaSN, this integral \detinG\
is evaluated by rewriting the determinant as a fermion integral
and then integrate out $M$. The resulting expression
is an effective theory on the single FZZT-brane
\eqn\detinsG{
\bra\det(x-M)\ket=\int_{-\infty}^\infty{ds\o\rt{2\pi g_s}}e^{-{1\o2g_s}s^2}
f_N(x+is).
}
Here we introduced the function $f_N(x)=x^N$.
We can easily see that
\detinsG\ is nothing but the integral representation of the Hermite 
polynomial, which in turn is the $N$-th orthogonal polynomial $P_N(x)$
of Gaussian measure
\eqn\detisHN{
\bra\det(x-M)\ket=P_N(x)=\lf({g_s\o2}\ri)^{N\o2}H_N\lf({x\o\rt{2g_s}}\ri).
}

It is interesting that $f_N(x)=x^N$ can be thought of as
a ``classical''
value of the determinant. Namely, $f_N(x)$ is the value of
$\det(x-M)$ evaluated at the minimum of the Gaussian potential $\Tr M^2$ 
\eqn\fxcalassical{
f_N(x)=\det(x-M_0)~,\quad M_0={\rm diag}(0,0,\cdots,0).
}
From this viewpoint, \detinsG\ defines an integral
transformation ${\cal Q}$ which maps 
the ``classical wavefunction'' $f_N(x)$ into
the ``quantum wavefunction'' $P_N(x)$ 
\eqn\fNtoPN{\eqalign{
&{\cal Q}:~~\det(x-M_0)\mapsto \bra\det(x-M)\ket \cr
&({\cal Q}f)(x)=\int_{-\infty}^\infty{ds\o\rt{2\pi g_s}}e^{-{1\o2g_s}s^2}
f(x+is).
}}
Note that this relation holds for arbitrary $N$,
therefore $f_n(x)=x^n$ is mapped to $P_n(x)$ for all $n$
\eqn\fntoPn{
{\cal Q}:f_n(x)\mapsto P_n(x).
}

We can generalize this picture to the multi-point correlators of
FZZT-branes. Following the similar procedure as above, the $K$-point function
of determinants is written as a $K\times K$ matrix model \MaldacenaSN
\eqn\KptinG{
\lf\bra\prod_{i=1}^K\det(x_i-M)\ri\ket=\int_{K\times K}dS\,
e^{-{1\o2g_s}\Tr S^2}\det(X+iS)^N,
}
where $X={\rm diag}(x_1,\cdots,x_K)$. This can be thought of as an effective
open string theory on the $K$ FZZT-branes.\foot{After taking a double scaling
limit, this open string theory on the FZZT-branes is identified as
the Kontsevich model \refs{\GaiottoYB,\MaldacenaSN}.}
After integrating out the angular part
of $S$, \KptinG\ becomes an integral over the $K$ eigenvalues of $S$
\eqn\fKmapGK{
\lap(x)\lf\bra\prod_{i=1}^K\det(x_i-M)\ri\ket
=\int\prod_{k=1}^K{ds_k\o\rt{2\pi g_s}}e^{-{1\o2g_s}s_k^2}
f^{(K)}(x_1+is_1,\cdots,x_K+is_K).
}
Here $\lap(x)=\prod_{i>j}(x_i-x_j)$ is the Vandermonde determinant
and $f^{(K)}(x)$ is given by
\eqn\fKandfN{
f^{(K)}(x_1,\cdots,x_K)=\lap(x)\prod_{i=1}^Kf_N(x_i).
}
As before, we regard $f^{(K)}(x)$ as a ``classical'' value of the
$K$-point function. The true correlator of determinants 
is given by the ${\cal Q}$-transform of $f^{(K)}(x)$ \fKmapGK.

To evaluate the integral \fKmapGK, we rewrite $f^{(K)}(x)$ \fKandfN\
as
\eqn\fKindetij{
f^{(K)}(x)=\det(x_i^{j-1})\prod_{i=1}^Kx_i^N=\det(x_i^{N+j-1})
=\det(f_{N+j-1}(x_i)).
}
Now the integral \fKmapGK\ is easily evaluated
by recalling that the integral transformation ${\cal Q}$ 
maps $f_n(x)$ into $P_n(x)$ \fntoPn. Therefore, the $K$-point correlator
of FZZT-brane is given by 
\eqn\multiinPN{
\lap(x)\lf\bra\prod_{i=1}^K\det(x_i-M)\ri\ket=\det(P_{N+j-1}(x_i)).
}
As expected, this agrees with the general formula of the
correlator of determinants \MorozovHH.
In \multiinPN\ we have multiplied the Vandermonde determinant $\lap(x)$
to the correlator of determinants. 
This factor makes 
the left hand side of \multiinPN\ anti-symmetric in $x_i$,
therefore FZZT-branes become fermionic \MaldacenaSN.
To summarize, we found the following
relation
\eqn\QmultiG{
{\cal Q}:~~\lap(x)\prod_{i=1}^K\det(x_i-M_0)~~\mapsto~~
\lap(x)\lf\bra\prod_{i=1}^K\det(x_i-M)\ri\ket.
}

In the next section, we will see that the transformation ${\cal Q}$
is exactly what we are looking for, {\it i.e.},
${\cal Q}$ maps A-brane amplitudes into B-brane amplitudes.

\newsec{D-brane Amplitudes on Conifold}

\subsec{A-branes in the Calabi-Yau Crystal Picture}
As shown in \refs{\OkounkovSP,\IqbalDS}, the target space theory of topological A-model
is reformulated as a certain $U(1)$ gauge theory and it leads
to the quantum foam picture of K\"{a}hler gravity,
which in turn is related to the statistical model of crystal melting.
In the case of resolved conifold, the relevant crystal is identified in
\OkudaMB. It is given by a lattice 
in positive octant of ${\Bbb R}^3$ with lattice spacing $g_s$,
and it is restricted to the interval $[0,Ng_s]$
in one direction. This statistical problem is solved by
slicing the 3D partition into a sequence of 2D partitions and
use the relation between 2D Young diagram and free fermion \OkounkovSP.

In this crystal picture, a non-compact A-brane is represented by
a half-line in ${\Bbb R}^3$ and it creates a defect in the lattice \SaulinaDA.
It is further shown that the presence of defect is represented
by the insertion of a certain fermion operator $\Psi_D$
in the free fermion computation mentioned above.
If the half line associated with the $i$-th D-brane 
ends on one of the axis of ${\Bbb R}^3_{+}$ at $a_i=g_s(N_i+\hf)$,
we identify the moduli $x_i$ of this D-brane as \foot{This definition
of $x_i$ defers from the usual convention by
a factor of $q^{\hf}$. Our definition is convenient
when comparing to the B-brane amplitude in the next section.}
\eqn\xitoNi{
x_i=q^{N_i}.
}
We are interested in the amplitude of Lagrangian A-brane with moduli $x$
\eqn\ZNasPsiD{
\CZ_N(x)=\bra \Psi_D(x)\ket.
}
This is first obtained by a Wilson loop calculation
in Chern-Simons theory on $S^3$ \OoguriBV,
and later it is reproduced by the melting crystal picture
\refs{\SaulinaDA,\OkudaMB,\HalmagyiVK}. 
We will consider an A-brane with the corresponding half-line
ending on the interval $[0,Ng_s]$. Then the A-brane has 
a topology $S^1\times {\Bbb R}^2$ \refs{\AganagicGS,\AganagicNX,\SaulinaDA}.
The explicit form of the single brane
amplitude is given by
\eqn\ZNx{
\CZ_N(x)=\prod_{n=1}^N(1-xq^n)={L(x,q)\o L(xQ,q)},
}
where $L(x,q)$ is the quantum dilogarithm \FaddeevRS
%\foot{This name
%comes from the fact that the first term in the genus expansion is the ordinary
%dilogarithm ${\rm Li}_2(x)$
%\eqn\logLexp{
%\log L(x,q)=-{1\o g_s}\sum_{n=0}^\infty{B_n\o n!}g_s^n{\rm Li}_{2-n}(x).
%}}
\eqn\qlog{
L(x,q)=\prod_{n=1}^\infty(1-xq^n).
}
Similarly, the amplitude in the presence of $K$ A-branes
is given by \refs{\SaulinaDA,\OkudaMB,\HalmagyiVK}
\eqn\Zmulti{
Z(x_1,\cdots,x_K)=\prod_{i<j}(1-x_ix_j^{-1})\prod_{i=1}^K\CZ_N(x_i)
=\lap(x)\prod_{i=1}^K\CZ_N(x_i)\prod_{i=1}^Kx_i^{1-i}.
}
Although it is a little bit ad hoc, it seems natural
to drop the last factor $\prod_ix_i^{1-i}$ in order to make the
multi-point function of D-branes anti-symmetric under
the interchange of $x_i$'s\foot{Perhaps this factor might be canceled
by carefully analyzing the zero-mode of boson appearing
in the bosonization of fermion.}.
This is motivated by the intuition that non-compact D-branes
are fermions.
Then the $K$-point function of D-brane operators
becomes 
\eqn\multipsiD{
\lf\bra\prod_{i=1}^K\Psi_D(x_i)\ri\ket=\lap(x)\prod_{i=1}^K\CZ_N(x_i).
}
One can immediately notice
the similarity of this expression to the ``classical'' correlation function
\fKandfN\ in the Gaussian matrix model. We would like to argue that
this is not just a coincidence.
We will see below that the ``quantum'' version of \multipsiD\
is given by the correlator of determinants in the Chern-Simons matrix model.

\subsec{B-branes in Chern-Simons Matrix Model}
In this subsection, we will consider the D-brane amplitudes 
in the Chern-Simons matrix model.
In \MarinoFK, it is observed that the partition function
of Chern-Simons theory is written as a matrix integral.
It is then realized that this Chern-Simons matrix model naturally
appears in the B-model mirror of resolved conifold \AganagicWV.

Let us recall the derivation of Chern-Simons matrix model.
Using the Weyl denominator formula, it is easy to
see that the partition function of
Chern-Simons theory \Zalprod\ is written as an integral
\eqn\CSinu{
Z_{CS}=\int\prod_{i=1}^Ndu_i\prod_{i<j}\lf(2\sinh{u_i-u_j\o2}\ri)^2
e^{-{1\o2g_s}\sum_iu_i^2}.
}
This is almost a matrix model integral, but the measure factor
is not the usual Vandermonde determinant. By the following 
change of variables \refs{\Walker,\TierzJJ}
\eqn\utom{
m_i=e^{u_i}q^{-N}
}
we can change the measure factor in \CSinu\
into the usual Vandermonde determinant
$\lap(m)^2$. Then $Z_{CS}$ becomes an $N\times N$
hermitian matrix model with log-squared potential
\eqn\CSmatM{
Z_{CS}=\int dMe^{-{1\o2g_s}\Tr(\log M)^2}.
}

Now let us consider the D-branes in this model.
From the general argument \AganagicQJ,
non-compact B-branes correspond to the insertion of determinant operators
in the matrix model. We define the B-brane operator with moduli $x$
as
\eqn\Bxop{
B(x)=(-1)^Nq^{N^2+\hf N}\det(x-M)=\det(q^{N+\hf}M-q^{N+\hf}x).
}
We have put a $q$-dependent prefactor for later convenience.
This is partly motivated by the relation \utom\ that
$\log(q^Nm_i)$ is the natural Gaussian variable $u_i$ in the original
integral \CSinu. The additional factor $q^{\hf}$
is related to our definition of D-brane moduli $x$ (see footnote 2).
Then the B-brane amplitude is given by
\eqn\SNasBexp{
\CS_N(x)=\bra B(x)\ket,
}
where the expectation value is taken in the Chern-Simons
matrix model \CSmatM. From the Heine's formula, the expectation value
of determinant is the $N$-th orthogonal polynomial.
In the case of log-normal measure \CSmatM, the orthogonal polynomial
is known as the
Stieltjes-Wigert polynomial \Szego.\foot{In 
\OkudaMB, a unitary matrix model with
measure $\vartheta_3(q,e^{i\varphi})$ is considered.
The orthogonal polynomials on $S^1$ associated with this weight
are known as the Rogers-Szeg\"{o} polynomials \Szego. They are related to
the Stieltjes-Wigert polynomials by a change of variable.}
With our normalization of $B(x)$ \Bxop, 
the $N$-th Stieltjes-Wigert polynomial
is given by
\eqn\SWpoly{
\CS_N(x)=\sum_{k=0}^N\lf[\matrix{N\cr k}\ri]q^{k^2+\hf k}(-x)^k.
}
Here $\lf[\matrix{N\cr k}\ri]$ denotes the $q$-binomial which is defined by
\eqn\qbinom{
\lf[\matrix{N\cr k}\ri]={(q)_N\o (q)_k(q)_{N-k}}~,\qquad
(q)_n=\prod_{j=1}^n(1-q^j).
}
We can also consider the multi-point function of B-branes.
From the general formula of correlator of determinants \MorozovHH,
the $K$-point function of B-branes reads 
\eqn\Bmulti{
\lap(x)\lf\bra\prod_{i=1}^KB(x_i)\ri\ket=\det(\CS_{N+j-1}(x_i)).
}

\subsec{A Map from A-brane Amplitudes to B-brane Amplitudes}
Now we consider the relation between the A-brane amplitude
$\CZ_N(x)=\bra\Psi_D(x)\ket$ in the crystal picture
and the mirror B-brane amplitude $\CS_N(x)=\bra B(x)\ket$ in the Chern-Simons
matrix model.
To see this relation, let us rewrite
$\CZ_N(x)$ given in the product form \ZNx\ into a summation form \Szego
\foot{This 
is a special case of the
$q$-binomial theorem \Andrews. 
}
\eqn\binom{
\CZ_N(x)=\sum_{k=0}^N\lf[\matrix{N\cr k}\ri]q^{\hf k(k+1)}
(-x)^k.
}
By comparing this expression with $\CS_N(x)$ in \SWpoly, 
one can immediately see that they are almost identical
except for the exponent of $q$.
This difference is taken care of by the following integral
\eqn\intsk{
\int_{-\infty}^\infty{ds\o\rt{2\pi g_s}}e^{-{1\o2g_s}s^2}e^{iks}
=q^{\hf k^2}.
}
Therefore, $\CS_N(x)$ and $\CZ_N(x)$ are related by
\eqn\SWvsZ{
\CS_N(x)=\int_{-\infty}^\infty{ds\o\rt{2\pi g_s}}e^{-{1\o2g_s}s^2}
\CZ_N(xe^{is}).
}
As advertised,  
this is nothing but the integral transformation
${\cal Q}$ appeared in the Gaussian matrix model! (Recall that
the variable obeying the Gaussian distribution is
$\log x$ in the Chern-Simons matrix model. Therefore, the transformation 
\SWvsZ\ is exactly the same as ${\cal Q}$ in \fNtoPN.)
To summarize, the A-brane amplitude and the B-brane amplitude
are related by the map ${\cal Q}$ 
\eqn\QZS{
{\cal Q}:\CZ_n(x)\mapsto \CS_n(x).
}

We can easily generalize this relation to the multi-point
correlators. Let us consider the 2-point function for illustration.
The 2-point function of A-branes is (see \multipsiD)
\eqn\twoA{
\lf\bra\prod_{i=1}^2\Psi_D(x_i)\ri\ket
=(x_2-x_1)\CZ_N(x_1)\CZ_N(x_2).
}
Using the recursion relation of $\CZ_N(x)$
\eqn\recursionZ{
xq^{N+1}\CZ_N(x)=\CZ_N(x)-\CZ_{N+1}(x),
}
\twoA\ is rewritten as
\eqn\twoANp{
\lf\bra\prod_{i=1}^2\Psi_D(x_i)\ri\ket=
q^{-N-1}\Big[\CZ_{N+1}(x_1)\CZ_N(x_2)-\CZ_N(x_1)\CZ_{N+1}(x_2)\Big].
}
By applying the ${\cal Q}$-map \QZS, this becomes the 2-point function
of B-branes \Bmulti, up to an overall factor $q^{-N-1}$.
One can easily generalize this argument for the general $K$-point
function, and it 
is found to be
\eqn\KptAindet{
\lf\bra\prod_{i=1}^K\Psi_D(x_i)\ri\ket=\det(\CZ_{N+j-1}(x_i)).
}
Here we suppressed an overall $q$ dependent factor for
simplicity. Clearly, 
\KptAindet\ is mapped to \Bmulti\ by the ${\cal Q}$-map
\QZS.  
In this way we find that the A-brane correlator
and the B-brane correlator are related by the ${\cal Q}$-map
\eqn\PsitoB{
\lap(x)\lf\bra\prod_{i=1}^KB(x_i)\ri\ket
=\int_{-\infty}^\infty\prod_{i=1}^K{ds_i\o\rt{2\pi g_s}}e^{-{1\o2g_s}s_i^2}
\lf\bra\prod_{j=1}^K\Psi_D(x_je^{is_j})\ri\ket.
}
Recalling the form of A-brane correlator \multipsiD,
this relation is written in a similar form as the Gaussian
matrix model case \QmultiG
\eqn\QABsche{
{\cal Q}:~\lap(x)\prod_{i=1}^K\CZ_N(x_i)~\mapsto ~
\lap(x)\lf\bra\prod_{i=1}^KB(x_i)\ri\ket.
}
Finally, if we introduce the ${\cal Q}$-transform ${\cal Q}\Psi_D$
of A-brane operator $\Psi_D$, \PsitoB\ can be written 
in a more suggestive form
\eqn\Qpsicorr{
\lap(x)\lf\bra\prod_{i=1}^KB(x_i)\ri\ket
=\lf\bra\prod_{i=1}^K{\cal Q}\Psi_D(x_i)\ri\ket.
}
Note that ${\cal Q}\Psi_D(x)$ 
is smeared by the Gaussian integral with width $g_s$,
thus it is no longer a local operator in 
the $x$-space. Perhaps it might be related to the noncommutativity
in spacetime.

\newsec{Discussion}
In this paper, we found that the A-brane correlators and the
B-brane correlators are related by the transformation ${\cal Q}$,
which is just a Gaussian integral.  
We would like to understand the physical meaning of this relation.
By the analogy to the Gaussian matrix model case,
it seems natural to regard the A-brane amplitude $\CZ_N(x)$
as a classical value of the determinant in the Chern-Simons matrix model.
From the product form of $\CZ_N(x)$ \ZNx, 
the classical background $M_0$ corresponding to
$\CZ_N(x)$ is given by
\eqn\Mzeroq{\eqalign{
\CZ_N(x)&=(-1)^Nq^{\hf N(N+1)}\det(x-M_0),\cr
 M_0&={\rm diag}(q^{-1},q^{-2},\cdots,q^{-N}).
}}
In other words, $\CZ_N(x)$ vanishes at 
\eqn\logxgs{
\log x=g_s,2g_s,\cdots, Ng_s,
}
{\it i.e.}, the zeros of $\CZ_N(x)$ are equally spaced.\foot{In 
\deHaroRZ, the eigenvalue density of Chern-Simons matrix model
is studied numerically at finite $N$. It is observed that
the density has $N$ peaks with almost equal spacing. This is consistent with
our identification of $M_0$ as the classical configuration of Chern-Simons 
matrix model.}
This is consistent with the picture that
the A-brane probes the background lattice of
the crystal. 
Namely, the A-brane amplitude obtained in the crystal picture
is a probe approximation ignoring the backreaction.
Moreover, it is known \refs{\AganagicQJ,\DijkgraafVP} that the A-brane amplitude 
is the zero energy wavefunction of the Hamiltonian $H(u,-g_s\del_u)$,
where $H(u,v)$ defines the mirror Riemann surface $H(u,v)=0$.
In our case, $\CZ_N(e^u)$ is the zero-mode of the Hamiltonian
\eqn\Huv{
H(u,v)=1-e^u-(1-e^uq^N)e^v.
}
This agrees with the known mirror of the resolved conifold
 \refs{\AganagicWV,\HoriKT} up to
a linear canonical transformation of the $u,v$ coordinates \AganagicQJ.

On the other hand, the B-brane amplitude $\CS_N(x)$
includes the effect of deformation of closed string background
due to the presence of D-brane \AganagicQJ. 
In fact,
the insertion of determinants is equivalent to a certain shift
of matrix model potential \refs{\AganagicQJ,\MaldacenaSN}
\eqn\dettopot{
\lf\bra\prod_i\det(x_i-M)\ri\ket=
{1\o Z_{CS}}\int dM\exp\lf[-{1\o2g_q}\Tr(\log M)^2+\sum_i\Tr\log(x_i-M)\ri].
}
As argued in \AganagicQJ, the deformation of matrix model
potential due to the insertion of B-branes
corresponds to turning on gravitational descendants of
K\"{a}hler class of ${\Bbb P}^1$.
It is remarkable that this deformation is captured by the simple
map ${\cal Q}$. 
Another way to see this effect is to look at the zeros of
$\CS_N(x)$. For example, $\CS_2(x)$ is factorized as 
\eqn\Stwo{\eqalign{
\CS_2(x)&=1-q^{3\o2}(1+q)x+q^5x^2=q^5(x-\al_+)(x-\al_-),\cr
\al_{\pm}&=\hf q^{-{7\o2}}\Big[1+q\pm\rt{(1-q)(1+3q)}\Big].
}}
The zeros are no longer equally spaced.
This might be seen as the distortion of crystal
due to the insertion of brane. One might expect that the ``quantum''
wavefunction $\CS_N(x)$ is significantly different from its classical
counterpart $\CZ_N(x)$. 
It would be interesting to study the large $N$ behavior
of $\CS_N(x)$. It would be also interesting to consider a double-scaling
limit of Chern-Simons matrix model as in the case of Gaussian matrix
model \MaldacenaSN.

In this paper we did not consider the insertion of anti-brane.
In the crystal picture, it is known how to construct the anti-brane
operator $\Psi_{\b{D}}$ \SaulinaDA. However it is not clear to us
what is the corresponding operator in the Chern-Simons
matrix model.
We leave this as a future problem.

\vskip 6mm
\noindent
{\bf Acknowledgment:} I would like to thank Marcos Marino
for correspondence.
\listrefs
\bye